# Teaching Design Science as a Method for Effective Research Development

Oscar Pastor, Mmatshuene Anna Segooa and Jose Ignacio Panach

**Abstract.** Applying Design Science Research (DSR) methodology is becoming a popular working resource for most Information Systems (IS) and Software engineering studies. The research and/or practical design problems that must be faced aim to answer the question of how to create or investigate an artifact in a given context. Precisely characterizing both artifact and context is essential for effective research development. While various design science guidelines and frameworks have been created by experts in IS engineering, emerging researchers and postgraduate students still find it challenging to apply this research methodology correctly. There is limited literature and materials that guide and support teaching novice researchers about the types of artifacts that can be developed to address a particular problem and decision-making in DSR. To address this gap in DSR, in this chapter, we explore DSR from an educational perspective, explaining both the concept of DSR and an effective method for teaching it. This chapter includes examples of DSR, a teaching methodology, learning objectives, and recommendations. Moreover, we have created a survey artifact intended to gather data on the experiences of design science users. The goal is to discover insights into contemporary issues and needs for DSR best practices (and, in consequence, evaluate the methodology we aim to teach). Our survey results offer a comprehensive overview of participants' experiences with DSR in the SE and IS Engineering domain, highlighting broad engagement primarily initiated during PhD studies and driven by supervisory guidance and research problems. We finally disclose the artifact to the community so that it can be used by other educators as a preparation when planning to teach DSR in tune with the experiences of their target audience.

Oscar Pastor, Research Center on Software Production Methods (PROS), Universitat Politècnica de València, Valencia, Spain, e-mail: opastor@dsic.upv.es

Mmatshuene Anna Segooa, Informatics Department, Tshwane University of Technology, Faculty of ICT, Pretoria, South Africa: segooama@tut.ac.za

Jose Ignacio Panach, Escola Tècnica Superior d'Enginyeria, Universitat de València, Valencia, Spain, e-mail: joigpana@uv.es

# 1 Introduction

DSR has evolved rapidly as a common research methodology in the Information Systems (IS) and Software Engineering (SE) disciplines, which is aimed at designing innovative and practical Information Technology (IT) artifacts, such as conceptual models and software systems [1], through diverse approaches that make use of IS/SE engineering approaches such as innovative platforms, architectures, and technologies for the engineering of any specific kind of IS.

Natural science helps humans understand the world around them and to seek explanations for the seemingly unexplainable. This is why IS studies frequently follow a natural science-oriented cycle, which suggests the structure that follows the steps of: problem definition, literature review, hypothesis development, data collection, analysis, results, and discussion [4]. This is directly connected with the SE life cycle to be followed for the design and implementation of software systems for the IS under study. Our work has always connected IS development and SE practice through the proposal of Conceptual Modeling Programming (CMP) approaches [13, 12, 3]. Following that working direction, in this chapter, we focus on a DSR application that can be applied to either IS and SE research working environments indistinctly.

According to Hevner [9], DSR is poised to take its rightful place as an equal companion to natural science research in the IS field. Therefore, IS scholars have realized the significance of DSR in improving the effectiveness and utility of IT artifacts when solving real-world business problems [8]. [15] defines design science as the design or investigation of an artifact in a context that may follow either an engineering cycle to design an artifact or an empirical cycle to investigate a knowledge problem.

In this chapter, we adhere to Wieringa's perspective on DSR, which provides a clear and explicit distinction in characterizing research works. The distinction lies in either focusing on designing an artifact intended to solve a particular stakeholder problem, thereby achieving their goals (leading to a design cycle) or addressing knowledge questions that require empirical research studies (leading to an empirical cycle).

This initial step in DSR characterization is crucial to delineate whether the research deals with a practical problem, the resolution of which will "change the world" by creating something that did not exist before (the designed artifact). Alternatively, the research may aim to answer a knowledge question, contributing to extending knowledge about a specific problem without "changing the world" to create a new concrete artifact. This distinction is vital because the working cycle for conducting a well-structured research project depends on this selection: practical problems versus knowledge questions.

Solving problems in a research-oriented manner needs to adhere to a design cycle strategy, while research aimed at addressing a knowledge question must follow an empirical cycle strategy. Since these two approaches involve different strategies, they also follow distinct steps. A design cycle encompasses problem investigation, treatment design, design validation, treatment implementation, and implementation evaluation. On the other hand, an empirical cycle involves an initial knowledge problem investigation, selection of the research design type, design validation, execution of the research (experiment), and evaluation of the results.

There is significant potential in advancing this domain by supporting emerging researchers with scientific materials that guide and teach them the application of DSR, considering its various dimensions, whether design-based or knowledge question-based, as a well-structured methodology. However, its application is not simple. Several significant decisions challenge the correct practice of DSR: Is the research problem centered around designing or improving an artifact (a practical problem), or is it, on the contrary, focused on answering knowledge questions? Is it a combination of both? In such cases, how does one integrate a design cycle, which involves creating an artifact, with its evaluation that requires result analysis to answer empirical research questions? Currently, our teaching experience in this context has identified gaps related to documenting the user experience to assess correct DSR practice. This serves as the central problem motivating our work.

All that leads to the problem of how to teach DSR correctly. What does it mean to apply DSR "correctly"? How do we identify the aspects that must be considered, and how do we evaluate the corresponding user experience? Moreover, how can these aspects be collected in a manner that can be used to gather practical data concerning DSR use? These problems have not received sufficient attention from DSR experts and are not well documented. Furthermore, there is a need for continuous assessment of DSR in IS and SE research. This chapter addresses also these problems through a study that aims to bridge the gap between DSR teaching and practice by presenting a survey for DSR assessment as a research methodology for IS and SE, which the DSR community may utilize for ongoing assessments.

The chapter's contribution is fourfold: First, it summarizes what DSR is, using an illustrative example. Second, it presents how to teach DRS, including learning objectives, thematic units, teaching methodology, teaching plan, and experience-based recommendations. Third, it proposes a concrete survey with the objective of gathering valuable data on the experience of DSR users to assess its practical use and discover good practice recommendations. Fourth, we report on selected results from running the survey from completed Master's and PhD theses at a university in Europe and a university in Africa.

This survey allows us to not only suggest further extensions of the DSR methodology, but to also clarify what aspects should be considered when teaching DSR in practice. The survey itself can be seen as a tool that, on the one hand, helps to better understand the particularities of DSR teaching, and, on the other hand, can be used also by others to generate a body of practical knowledge about how to improve design science-based teaching.

To achieve our objectives, we maintain consistency with our interest in DSR by following a Wieringa-based approach [15]. We begin by summarizing the DSR main characteristics and fundamentals in a Nutshell (section 2), before discussing how to teach DSR (section 3). Next, we state the problem and its need (problem investigation, section 4), and introduce the survey artifact to analyze how DSR is perceived by students (section 5). Subsequently, we report on the results and discuss them in section 6, before concluding our chapter in section 7.

## 2   Design Science Research: Fundamentals in a "Nutshell"

DSR can be defined as the design and investigation of artifacts in context [15]. The **artifacts** are designed to interact with a **problem context** in order to improve something in that context. The artifact itself does not solve any problem. It is the interaction between the artifact and a problem context that contributes to solving a problem. An artifact may interact differently with different problem contexts and solve different problems in different contexts. Examples of artifacts are methods, services, tools, business processes, or software systems, while examples of contexts are: norms, desires, people, or organizations.

DSR is divided into two parts: **Design problems** and **Knowledge questions**. Design problems call for a change in the real world. A solution is a design, and there are usually many different solutions. Knowledge questions do not call for a change in the world but ask for knowledge about the world as it is. To see examples, let's use the title of a research work; that must be its best summary and must help to identify both artifact and context well.

For an example of a design problem, let's take this research work's title: "Design of a usability and usefulness test to evaluate the quality of cloud service providers." The artifact is the test itself, while the context is the quality evaluation of cloud service providers. An example of a knowledge question would refer to a research work intended to find out if the test is really usable and useful for cloud service providers, having as title "Evaluation of the usability and usefulness of a test to assess cloud service providers quality."

It is not always so simple to identify and distinguish artifacts and contexts. Look at this other example: imagine a research work with the title "From Adoption to Endurance: Exploring the Dynamics of AI Adoption Across Time and Contexts". While context is well-identified (indeed, the whole title appears to introduce the context), what is the artifact in this case? How the context under analysis is going to create one solution? What solution? This is a good example of how teaching practices must take this crucial point as a central one for discussing good DSR practice and recommending research titles where both artifact and context are clearly identified, making true the statement that the research title must be the best summary of a research work.

Two different dimensions help to characterize the context correctly: social and knowledge. The problem context of an artifact can be extended to the stakeholders of the artifact and with the knowledge used to design the artifact. This extended context is the context of the DSR project as a whole. The resulting picture is a framework for design science, shown in Fig. 1, that should help to identify the context of a research work. The **social context** determines who are the stakeholders that will benefit from the use of the artifact, what their goals are, how they will be accomplished, or what are the funding constraints of a research project. The **knowledge context** consists of existing theories from science and engineering, specifications of currently known designs, valuable facts about currently available products, lessons learned from the experience of researchers in earlier design science projects, and plain common sense. DSR uses this knowledge and may add to it by producing new designs or answering knowledge questions.

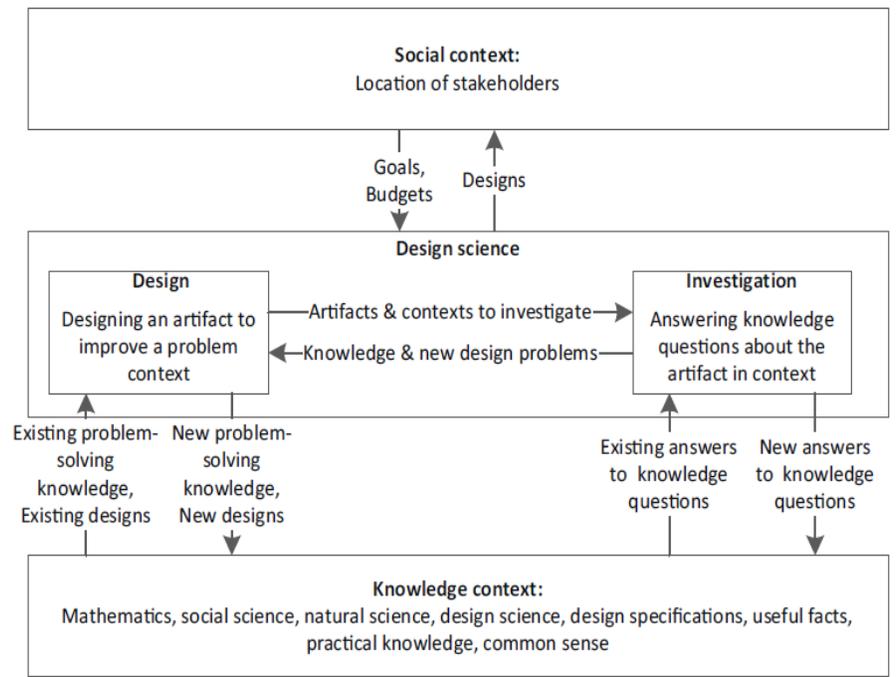

**Fig. 1** A framework from Design Science [15].

For example, in the social context of genomic data analysis, we aim to design an artifact to help data analysts interpret the genomic information correctly. So, the investigation to answer is, "Is the system's usability enough to help improve the genomic data interpretation?" In the knowledge context we find usability guidelines, templates of graphical user interfaces, and design recommendations.

DSR is conducted in a design cycle or in an empirical cycle. The **design cycle** is associated with a Design problem. It includes the definition of the problem investigation, the treatment design, and the treatment validation. The result of the design cycle is a validated treatment ready to be transferred to the real world. Depending on the research strategy, the design cycle can be divided into different subcycles in order to have a large cycle that gathers all the research. Each subcycle must specify the different tasks to be fulfilled in order to reach the goals. For instance, this can happen if the research includes a sequence of component designs, and each one requires an individual design (sub)cycle. Or if each component of the design cycle has itself a cycle associated to its development.

Fig. 2 shows an example of a design cycle divided into three subcycles for this specific illustration. This example is the one previously described to design an artifact to help data analysts to interpret genomic information. The tasks of each cycle are represented with the prefix "T" in Fig. 2. The first subcycle affects the problem investigation step. It includes tasks for studying the current state of the problem, the social context (or stakeholders), and the comparison of the results with existing works. The second subcycle is the treatment design, which aims to define artifacts in order to conduct the investigation. This cycle defines the knowledge context (or theories) of the research. The third subcycle is treatment validation, where empirical studies are conducted to analyze whether or not the results meet expectations.

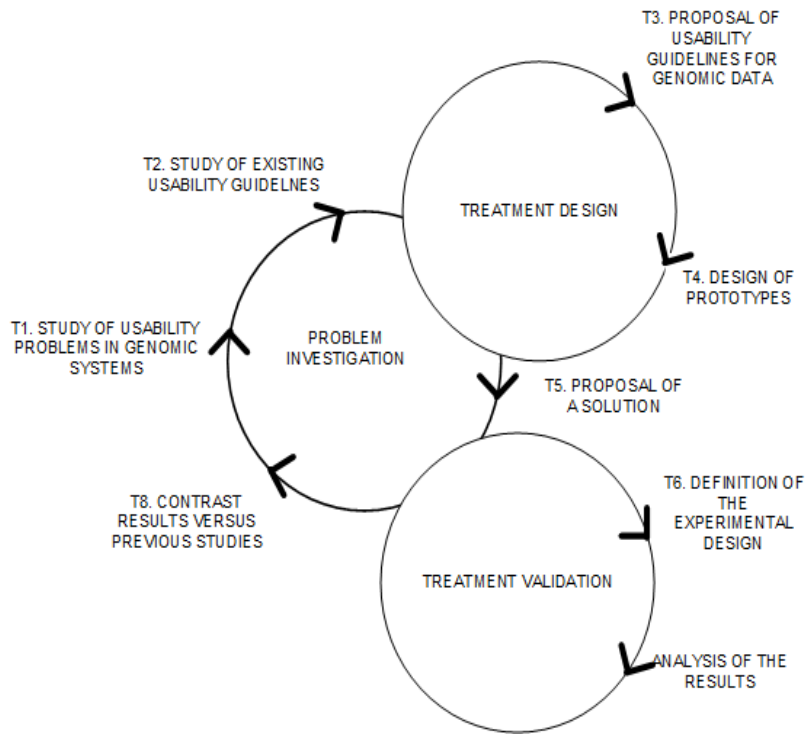

**Fig. 2** Example of a design cycle to build usable systems in genomic data analysis.

The empirical cycle is associated with a knowledge question, and it is structured as a checklist of issues to decide when a researcher designs a research setup and wants to reason about the data produced by this setup. The cycle starts with a list of questions about framing the research problem. The rest of the checklist is about designing the research setup and the inferences from it, about research execution, and about data analysis.

The checklist of the empirical cycle is a logical grouping of questions that help you to find justifiable answers to scientific knowledge questions. The checklist depends on the specific research. An example of a checklist may include: the motivation of the experiment, the research questions, the definition of variables, the specification of metrics, the definition of the experimental design, the definition of the protocol, and the mitigation of threats to validity.

This cycle aims to answer knowledge questions that cannot be answered using the literature, asking experts or testing a prototype. The cycle includes tasks to define the research problem, design the research setup, research execution, and data analysis. In the same way, as in the design cycle, the empirical cycle can be divided into subcycles.

Fig. 3 shows an example of an empirical cycle to analyze whether or not usability guidelines improve the usability of the graphical user interfaces and what guidelines have more effect on the user experience. Note that in this example, the research does not propose a new approach; it aims to evaluate existing approaches.

The example is composed of three subcycles, each one with specific tasks identified in Fig.3 with the prefix "T". The first subcycle aims to define the context, which includes the definition of the research questions, the required profiles of the population, the experimental design, and the generalization of results once the experiment is over.

The second subcycle is the experiment execution, where we recruit the population and we conduct the experiment to get the measurement. The third subcycle analyses the results with statistics and interprets such data.

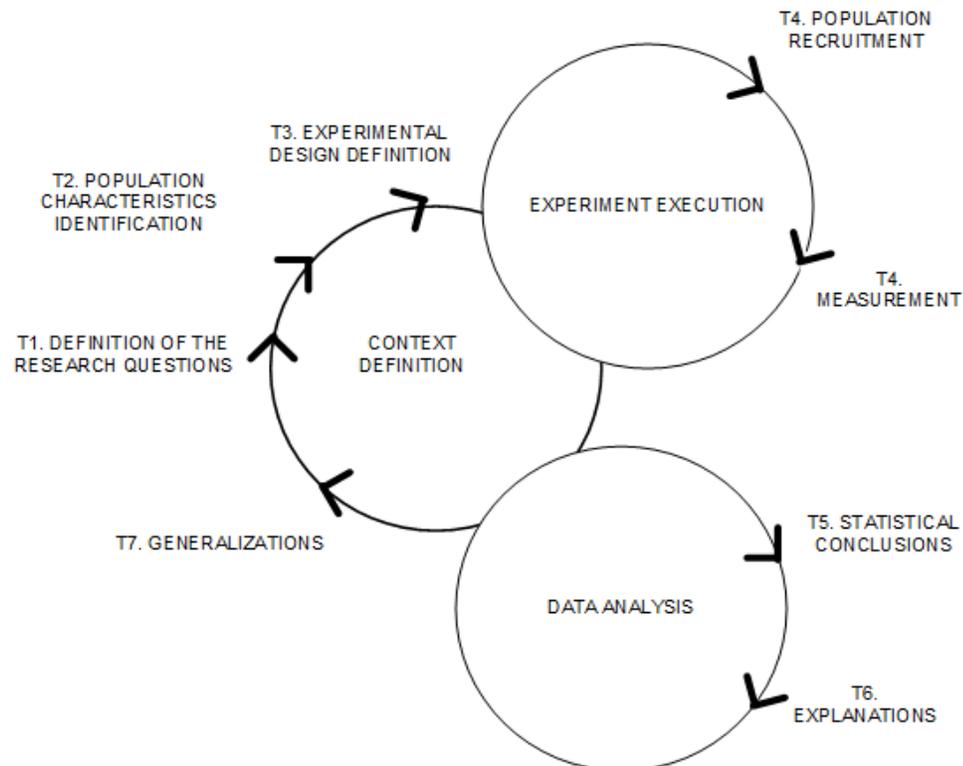

**Fig. 3** Example of an empirical cycle to evaluate usability guidelines.

Summarising, four main aspects are to be considered relevant for a sound DSR teaching process which educators may take into consideration when formulating DSR learning outcomes:
1. The first aspect to take into account is how to identify the precise artifact that is designed and the proper context in which it is intended to work. The correct identification of both artifact and context becomes an essential initial step of any DSR process, as we have already discussed earlier.

2. Another aspect to consider is selecting an adequate title, where both the artifact and the context can be clearly recognized.
3. A third aspect is to delimit the scientific nature of the research. Does the research core face a design problem or a knowledge question? Is there a new artifact as a result of the research work that did not exist before? Or, on the contrary, the research work is about accumulating knowledge concerning already existing artifacts?
4. Once the research nature has been characterized, the fourth, and last aspect refers to selecting the particular research cycle (design cycle or empirical cycle) that better fits the planned research work that must be developed.

These four aspects and considerations of how DSR is used in practice, together with all the dimensions discussed in the previous sections, have guided the learning objectives and thematic units that should structure a basic course to teach adequately Design Science. The next section describes how to teach DSR, covering both the theoretical foundations and the practical skills necessary to engage in this innovative research methodology.

## 3 How to Teach Design Science Research

This section describes how to teach DSR in a potential course for researchers in IS and SE, especially for PhD students who are starting their research. The recommendations shared in this section are rooted in our experiences of teaching DSR in our own educational environments and tutorials given in the research community, and they should be also understood as such experience-based recommendations. We elaborate on the course's objectives, how it is divided into thematic units, the teaching method, the teaching plan, followed by practical recommendations and advice. Classes should combine theory and practical exercises, ensuring that all theoretical concepts can be seen in real-world applications.

We begin by outlining the learning objectives that students are expected to achieve upon completing a course:

- **Objective 1:** Students should be able to define and explain key concepts of design science.
- **Objective 2**: Students should demonstrate the ability to apply design science principles to real-world problems.
- **Objective 3:** Students should effectively work in teams, contributing to and learning from their colleagues.
- **Objective 4:** Students should exhibit strong critical thinking skills, able to evaluate their work and make improvements based on feedback.
- **Objective 5:** Students should be able to integrate design science principles into their PhD or MSc research, producing high-quality, rigorous academic work.

In order to reach such objectives, the course is divided into **thematic units**. Thematic units have been extracted from the book written by Wieringa [15], which describes how to apply DS to computer science in detail.

- **Unit 1:** What is design science?
- **Unit 2:** Research goals and research questions
- **Unit 3:** The design cycle
- **Unit 4:** The empirical cycle

These thematic units are taught using a teaching method that begins with the problem definition. This involves identifying a problem or need within the realm of SE, such as developing new software, optimizing an existing system, or creating new software development method, among others. This problem will be used throughout the entire teaching process. The teacher defines the problem on their own, and it is recommended to choose a problem with which most of the students are familiar. Moreover, the teacher divides the students into several groups of 3 or 4 students. These groups will participate in the following steps of the teaching method.

Once the problem has been defined, the method involves describing all the elements in a theoretical class. All the elements shown in Fig. 1 are explained. Each concept is defined as described in the definition of the DSR method. Next, using the previously defined problem, each group of students must think about how the concept can be operationalized for that specific problem. For instance, this is the right time to determine if the research a centered around a practical problem, a knowledge question, or both. A precise answer to these questions will facilitate the correct DSR application process. First, there is an initial discussion within each group. After 5 minutes of internal discussion, each group shares their thoughts with the other groups. Each concept involves brief brainstorming within the groups and among the groups, facilitating the generation of ideas. At the end, the teacher decides which ideas are the most suitable for that problem.

After introducing all the elements of the DSR method, and once the nature of the research is delimited (practical problem, knowledge question, and their potential connections) the next step is to consider the necessary cycles that depend on the research nature decision. Each group must determine whether the research method requires one or more cycles, which of these cycles are to define the approach (design cycle), and which ones are required to validate those approaches (empirical cycle). Additionally, each team must specify the different tasks required in each cycle. After one hour of group work, each group shares their conclusions with the other groups. The teacher decides which proposals are the most suitable for that specific problem, discussing the pros and cons of each solution. Once the concepts of the DSR method have been introduced, and students know how to define the different cycles, the final part of the teaching method consists of applying the DS method to each student's PhD thesis. Groups are eliminated, and each student works individually. Usually, the DSR course is taken by PhD students, so each student has a different research project where DSR can be applied. From now on, we assume that the audience of the proposed course are PhD students, although the ideas can be applied to any academic research working context.

Working individually, each student must specify how the different terms of DSR can be applied to their research, as well as think about the necessary cycles and tasks. A short document must be prepared to show specifically how DSR can be applied. Elaborated documents are reviewed by the teacher outside the classroom. Written feedback can be provided to each student in such a way that, beyond learning DSR, the result can be included in the PhD document as part of the student's work.

The teaching plan should span at least 4 hours to be able to teach the minimum while it can be gradually expanded. In that sense, we propose at least two lecture units or a half-day tutorial to teach the essence of design science. Effectively teaching DSR can as well be captured in an entire course, that could emphasize particular aspects of the practical application of Design Science depending on the setting and which learning objectives the educator might want to stress. Next, we describe the **teaching plan** considering the minimum time we recommend to spend on key aspects (4 hours in total). In that sense, a course can be also effectively applied as part of tutorials and adopted to the corresponding formats. Below is the timeline:

- **From 0 minutes to 15 minutes**: The teacher presents the problem to be discussed during the class and creates the groups of students.
- **From 15 minutes to 1 hour and 30 minutes**: The teacher presents the different

elements of DS. Each element is worked on within each group and then shared among the other groups.
- **From 1 hour and 30 minutes to 2 hours and 30 minutes**: Each group must specify the different cycles to solve the described problem. There is a discussion first within each group and then among the groups.
- **From 2 hours and 30 minutes to 4 hours**: Each student must individually apply DSR to their PhD research and elaborate a short document with the results.

Finally, we summarize some **recommendations** both, from the point of view of the teacher and from the point of view of the student. From the point of view of the teacher:

- **Choose real problems**: Select problems that are relevant and familiar to the students. This enhances engagement and makes the learning process more meaningful.
- **Group work**: Encourage collaborative learning by organizing students into groups. This fosters teamwork and allows for the exchange of diverse ideas. The results of the group work do not need to be perfect but should serve as a way to discuss several alternatives.
- **Regular feedback**: Provide timely and constructive feedback throughout the learning process to guide students and help them improve their work. It is recommended to provide written feedback individually to each PhD thesis. This way, all details can be processed calmly by the student.
- **Problem used as a guide throughout the learning process**: Choose a problem that is easy enough for all students to understand the context and the problem itself. It is not advisable to start with a problem related to a real PhD thesis, as these topics are very specific and not all students may understand them. Try to adjust the problem to the audience.
- **Encourage critical thinking**: To properly apply DSR to each PhD research, the student must have learned DSR properly beforehand. The teaching process should strongly involve discussion both within the groups and among the groups. If the students do not participate actively in the discussions, the teacher must encourage and facilitate this discussion. It is crucial that students learn from the discussion what is correct and what is not. In classes with poor discussion, more errors are observed when students apply DSR to their PhD research.

From the point of view of the students, these are the recommendations:

- **Participate actively in the group work**: Collaborate with your peers in group activities. This fosters teamwork, allows for the exchange of diverse ideas, and helps you learn from different perspectives. Remember, the goal is not perfection but to explore various alternatives and learn from the process.
- **Seek and reflect on feedback**: Regularly seek feedback from the teacher and colleagues of the group. Constructive feedback is crucial for your improvement. This can for instance help to clearly identify what is provided by the research (concrete artifact) and what is the context of interest for such a solution, which are crucial dimensions for conceptually characterising a research contribution. When the feedback of the PhD research is provided at the end of the course, take time to reflect on it. DSR properly applied to your research is an essential section in your PhD memory.
- **Engage in discussions**: Actively participate in discussions during class. Whether in group discussions or class-wide debates, sharing your thoughts and listening to others will deepen your understanding. Do not be afraid to speak up, even if you are unsure, these discussions are valuable learning opportunities.

- **Document your process**: Keep detailed records of your design process. Documenting your steps, decisions, and reflections will help you track your progress and provide valuable insights for future projects. Taking notes and ideas in the discussion stage can be very helpful when you apply DS to your specific research. Words easily go with the wind.

## 4 Problem Investigation

Despite the fact that more than a decade of DSR discussions and the publication of the seminal articles by Gregor, et al [6], Hevner et al [10], Kuechler. & Vaishnavi [11], Peffer et al [14] and Wieringa [15] which led to significant growth in DSR popularity in IS and SE, good DSR practice is still a problem for researchers interested in using DS as their research methodology. Literature highlights the emerging development of DSR methodology in postgraduate studies projects, which presents a challenge to understanding their research philosophy and the application of this methodology [2].

The reality is, however, that the experience of emerging researchers, such as doctoral students, regarding DSR adoption and application is still not well-documented [5, 2]. This renders the design of appropriate education and training cumbersome. Furthermore, there is limited support and guidance for documenting and effectively managing DSR processes. This indicates that the existing literature and materials that guide novice researchers on the underlying philosophy of DSR studies, the development of artifacts to address specific problems and decision-making on DSR guidelines to address issues need to have robust and well-documented support.

Given this shared background, it is evident that evaluating the application of DSR to improve the user experience and support the methodology constitutes a challenging journey. This justifies the development of a standard survey artifact that may serve as a tool for DS Custodians such as [8, 11, 14, 18, 20] and IS and SE engineering researchers in their ongoing work. The survey should include a set of questions that conceptually characterize the dimensions to be considered if we want to understand DSR practice. We want to improve how to teach it by identifying which aspects the teaching process should focus on, and we aim to collect data for further analysis.

The data collected using our survey artifact may contribute theoretically to the DSR literature gap (conceptual characterization of DSR methodology versus its successful practical application) by providing insights into the DSR user experience and the need for improvements in this domain. Lastly, the analysis results can also aid in assessing the maturity of the DSR methodology in solving real-life problems to deliver valuable artifacts in concrete contexts, providing value in a clear, well-delimited, and convincing way. All these objectives guide our work, leading to the design of the survey that we introduce in this work.

## 5 Survey Artifact for DSR Users' Experience

There is no guide to applying DSR correctly, and the practical application of some elements of the DSR teaching process (which we discuss in this chapter) can be confusing. To develop a solution to this practical DSR problem, we propose a version of a survey artifact that intends to gather data on design science users' experience to assess the methodology application, and discover insights on the contemporary issues and needs for DSR best practice. It should go without saying that this artifact may be further revised in the future to feet the individual needs of other scholars. In that sense, our survey may be re-used and adopted by other educators as a preparation for designing/aligning the course with the target audience.

**Table 1** Survey artifact for DSR application and experience including a first set of yes/no, multiple choice" and "open-ended" questions.

> 1. Have you used Design Science research prior PhD?
> 2. Where did you use design science methodology?
> 3. What motivated the research to take DSR approach?
> 4. What problem was the research solving? Or what was the research problem?
> 5. What type of artifact has the study produced?
> 6. Whose design science guidelines did your study follow?
> 7. Were you aware of other guidelines than the one you used for your study?
> 8. If yes, why did they go for the one you chose, if you selected no then type N/A
> 9. Did the main purpose of your study follow the empirical research cycle and/or engineering cycle?
> 10. Did you evaluate the artifact?
> 11. How did you evaluate the artifact?
> 12. Did you attend a Design science workshop before/during the study?
> 13. What challenges have you experienced when using design science methodology?
> 14. What opportunities have you experienced when using DSR methodology?
> 15. What are the lessons learned in the journey?
> 16. What kind of support do you wish you had received during the journey?
> 17. Were they ever discouragement from carrying on with the methodology during the study?
> 18. What is interesting about the methodology?
> 19. What is less interesting about the methodology?
> 20. Have you published your research conducted using design science methodology?

The survey draft was informed by the literature on the DSR research gap and the need to scientifically document DSR user experience especially for the case of postgraduate students with the aim to discover insights that will help DSR custodians and scholars in the IS Engineering and SE domain to provide the necessary support to apply best practice towards the methodology. Our survey, in its basic version, has 20 questions as indicated in Table 1. These questions are clustered into six (6) groups, whose conceptual justification is now introduced.

- Cluster 1: Experience: questions 1 to 2 focus on identifying the level of DSR experience as a scholar in the domain.
- Cluster 2: Purpose: questions 3 to 5 delimit the problem concerning the artifact intended to design (artifact identification-oriented).
- Cluster 3: DSR guidelines selection: questions 6 to 9 determine the DSR guidelines selected to accomplish the artifact's design in its working context (context characterization-oriented).
- Cluster 4: Evaluation: questions 10 and 11 explore if the artifact has been evaluated, and how.
- Cluster 5: User's experience: questions 12 to 19 investigate the support, experience, and satisfaction in the application of the DSR guidelines.
- Cluster 6: Dissemination: question 20 determines if the results of the project delivered through DSR have been published, emphasizing the sharing of information about DSR use and value.

## 6 A First Qualitative Analysis of the Preliminary Survey for DSR Users' Experience

In our attempt to employ the proposed survey to collect data that could help us to better understand and improve teaching design science, this section presents preliminary results obtained from the qualitative analysis of the survey. It is reported according to the six clusters that structure it. The analysis was conducted on a small sample size, as it was intended to pilot the artifact in preparation for data collection from a larger sample.

Our sample included participants who utilised DSR in their postgraduate studies, who are in scope of educational measures taken in order to prepare them for applying DSR.

Responses were received from 21 participants (4 doctoral, 14 master's graduates, and 3 bachelor's degrees). Feedback is useful for analyzing how DSR is applied in various research fields. Moreover, the results provide insights into the benefits and drawbacks of using DSR in a teaching context and indicate how to proceed to continuously improve such a DSR teaching experience.

## 6.1 Artifact Results and Discussion

This section introduces and discusses cluster by cluster on what the survey results meant in terms of the better understanding and continuous improvement of the teaching methodology that is to be followed for DSR teaching practice. In responding to the questions in cluster 1, which aimed to gather the DSR experience of the participants as scholars in the IS Engineering domain, two questions were posed. These questions were designed to identify users' experience and knowledge of DSR. Q1 inquired about prior DSR use (before either a PhD Thesis or an MSc Thesis), while Q2 focused on where participants had utilized design science methodology before.

The participants indicated that they have used DSR of which most of them highlighted that they used the methodology before their PhD studies. We conclude that it is important to know the DSR background of DSR users before starting a sound DSR teaching strategy, to calibrate well how detailed the introductory material must be.

Cluster 2, which aimed to define the problem that the artifacts were to address, was faced using three questions identified as Q3 to Q5, that focus on the survey instrument. Q3 inquired about the motivation behind choosing the DSR approach; Q4 sought to determine the problem the research aimed to solve or the research problem itself; and Q5 was designed to explain the type of artifact the study intended to generate.

The findings highlight that the motivation to conduct the study through design science came from the supervisor/study advisors while some participants followed the methodology to better characterize the problem that the study intended to address. The artifact that most participants produced was a model, system, or method.

In this context, an accurate artifact identification becomes an essential actor of the DSR teaching process. This confirms our practical experience, where the precise identification of the artifact that a research work is designing or informing, is what takes more time and stronger discussions. A research work must be built on a clear and concrete characterization of the artifact that is being designed.

Problems which were solved by the DSR studies included.
- Genomic data quality measurement.
- Design, develop, and validation of a platform for managing genomic data.
- Conceptual Model of the Genome to improve understanding of the genomic domain.

- Creating artifacts to teach systematic literature review.
- What model will inform the adoption of a smart learning environment for a Secondary School in Ekurhuleni North district?
- Lack of appropriate models to inform data governance relating to Big Data.
- Access to education in low bandwidth areas.
- Online fraud detection model.
- How can a Pharma company integrate Gen AI to stay competitive and innovative?
- Developing a method for identifying KPIs for a strategic alignment of IT Projects.
- How to diminish the impact of senior consultant turnover on knowledge management in a software editing company?
- What strategies are most effective in managing organizational change when transitioning to electronic invoicing in B2B processes?
- Rethinking/redesigning the Requirements Engineering process of an IS Department.
- What solutions can be implemented in a company to reduce the costs of IS?
- How can organizations implement strategies to increase user adoption of SSBI tools?

For better understanding when we face a design (practical) problem or a knowledge question, with their corresponding engineering cycle (in the design problem case) or the empirical cycle (for the knowledge question case), Table 2 shows the list of DSR problems that we have presented above, classified according to these characteristics.

From the perspective of the survey's assessment, it was positive to observe that very different topics aligned well with the survey's purpose, and that identifying the type of artifact at the core of the analysed research became a clear and informative task. The problem identified by the survey reflects interest in both the design cycle and the empirical cycle. Some topics focused on design (practical) problems, where models and methods were developed. In contrast, others focused on knowledge problems, involving information-based investigation into how organizations can better implement strategies and studies on how to diminish the impact of senior consultant turnover. Our last conclusion in this context was how important it is to delimit clearly if the DSR process involves a practical problem or a knowledge question. Again, this confirms our practical experience around how to determine this aspect after having precisely identified artifact and context: artifact and context require a different treatment depending on being facing a practical problem or a knowledge question, as it has been discussed before in this chapter.

Cluster 3, where DSR guidelines were selected, aimed to determine the guidelines used to accomplish the artifact's design. In this cluster, four questions were posed on questions 6 to 9 of the survey instrument. Q6 asked about which DSR guidelines were followed in the study; Q7 inquired whether participants were aware of different DSR guidelines than the ones they used for their study; Q8 complemented the previous question, clarifying why the participant selected the chosen guidelines when the answer to Q7 was yes. Finally, Q9 asked about the main purpose of the study when following either an empirical research cycle and/or an engineering research cycle.

The results indicate that participants commonly used DSR guidelines from Hevner et al [10], Vaishnavi & Kuechler [11], Peffers et al [14], Wieringa [15]. Those who used Wieringa [15] tended to follow the engineering cycle, while participants who used Hevner et al [10] and Peffers et al [14] followed the Empirical cycle. Moreover, those who followed Wieringa [15] were not aware of other guidelines, while those who used Hevner et al [10], Peffers et al [14] and Vaishnavi

& Kuechler [11] were knowledgeable about alternative DSR guidelines.

Participants noted the reason for choosing Hevner et al [10] as its ease of understanding and support for model development. On the other hand, the choice of Peffers et al [14]. was attributed to the clarity and detail of its guidelines, along with the recommendation by supervisors.

**Table 2** Examples of DSR problems and cycles.

| Examples of DSR problems | DSR cycle | Type of DSR problem |
|---|---|---|
| Genomic data quality measurement. | Engineering | Design |
| Design, develop, and validation of a platform for managing genomic data. | Engineering | Design |
| Conceptual Model of the Genome to improve understanding of the genomic domain. | Engineering | Design |
| Creating artifacts to teach systematic literature review. | Empirical | Knowledge |
| What model will inform the adoption of a smart learning environment for a Secondary School in Ekurhuleni North district? | Empirical | Knowledge |
| Lack of appropriate models to inform data governance relating to Big Data. | Empirical | Knowledge |
| Access to education in low bandwidth areas. | Empirical | Knowledge |
| Online fraud detection model. | Empirical | Knowledge |
| How can a Pharma company integrate Gen AI to stay competitive and innovative? | Empirical | Knowledge |
| Developing a method for identifying KPIs for a strategic alignment of IT Projects. | Engineering | Design |
| How to diminish the impact of senior consultant turnover on knowledge management in a software editing company? | Empirical | Knowledge |
| What strategies are most effective in managing organizational change when transitioning to electronic invoicing in B2B processes? | Empirical | Knowledge |
| Rethinking/redesigning the Requirements Engineering process of an IS Department. | Empirical | Knowledge |
| What solutions can be implemented in a company to reduce the costs of IS? | Empirical | Knowledge |
| How can organizations implement strategies to increase user adoption of SSBI tools? | Engineering | Design |

Based on the questions asked in cluster 3 and the responses from the participants, we found that answering the questions became an easy task, and the provided information was useful for better understanding and comparative analysis.

We assume that DSR can be successfully applied using any of the existing DS approaches, and our conclusion here is that it is important to make DSR users be aware of the fact that different valid DSR approaches exist and that it is important to make explicit and known their strong conceptual connections. It is also fruitful to explain why a particular selection is done to be used in practice.

Cluster 4, focused on the evaluation of the artifact, aimed to establish if the artifact was implemented and how the evaluation was conducted. To address this cluster, two questions were presented through Q10 and Q11 of the survey instrument. Q10 inquired about whether the research work included artifact evaluation, while Q11 focused on how the artifact had been evaluated. From the findings, all the participants indicated that their artifacts were evaluated. The evaluation was done through experiments, surveys, focus groups, and interviews. It appears that all who used the engineering cycle of the DSR used experiments and interviews to evaluate the artifacts of their study, while those participants who followed the empirical cycle used surveys and focus groups.

Based on the questions asked in cluster 4 and the responses from the participants,

we also found the results informative, aiding in obtaining a better understanding of what was done and how, particularly in terms of distinguishing the practical problem vs. knowledge question dichotomy that we discussed earlier. The main implication of this distinction is that a design cycle focuses on the design of an artifact, and an implementation directly suggests the use of such a practical problem-based working context. In this case, the evaluation corresponds to a validation step, and it is then a part of the contribution.

Alternatively, an empirical cycle indicates that there is no implementation, and the value of the research concentrates on a strong experimental task that constitutes the core of the research contribution. An in-depth analysis of particular survey answers can allow to assess that these connections exist in order to confirm that a sound DSR practice is applied. On the other hand, accumulating data from many individual cases could make it possible to perform a data analytics process to confirm that correct DSR patterns appear when we compare a significantly diverse set of concrete Design Science applications cases.

Cluster 5 aimed to gather users' DSR application experience to investigate the support, experience, and satisfaction in the application of DSR guidelines. It covers the user's experience dimension. This cluster was addressed through eight questions on the survey artifact, specifically questions 12 to 19. These questions cover various aspects in this regard. Q12 explores whether participants attended any Design Science workshops before or during the study; Q13 asks about challenges experienced by DSR users when using the design science methodology; Q14 seeks declarations of opportunities obtained as a result of DSR methodology use; Q15's purpose is to find out lessons learned in the DSR application journey; Q16 detects any potential support that was lacking during the process; Q17 asks for potential discouragement from continuing with the methodology during its application; Q18 emphasizes aspects considered as more interesting by participants and that provide a concrete, useful benefit; finally, Q19, on the contrary, explores what was seen as less interesting about the methodology.

This cluster provided interesting information that showed us how collecting this type of info could help us better understand the particular context of DSR practice, what affects the planning of a DSR teaching process. In particular, the participants reported to have received training (except for one participant who indicated that they did not receive training during their studies). Understanding the design cycles seemed to be a challenge for the participants who did not receive training, emphasizing that the methodology is highly encouraged. Furthermore, other participants indicated that they experienced challenges in positioning the research problem into either design or empirical problems and selecting the evaluation framework. This mimics our practical experience, where the characterization of the research work as either a research (practical) problem or knowledge (empirical) question is always a complex task for students.

Therefore, a suggestion to provide more examples of the methodology is recommended to support students in understanding DSR. Additionally, training should be offered at the early stages of the study. Educators can also design the learning material based on the participant's needs and clear learning objectives and outcomes may be set to address these challenges.

Some participants highlighted that they were initially afraid to use DSR; however, they realized it was doable as they persevered and managed to complete their studies. Overall, the participants indicated they were not discouraged from using DSR, although some mentioned feeling discouraged along the journey as they engaged with the methodology.

This cluster facilitated the identification of methodological problems encountered by the participants in practice. The information was more diverse, yet informative, covering an aspect—methodological usability and the practical ease of

use—whose positive properties were not as clear and convincing as expected.
Cluster 6 aimed to establish if the work had been disseminated. Specifically, question 20 determined if the project's results delivered through DSR had been published, emphasizing the sharing of information about DSR use and value. To address this cluster, the question was whether participants have published the research conducted using design science methodology to determine if the DSR use was reported and documented. Out of 21 participants, only 19% (four participants) have published their work, some published in conference proceedings while other participants published their work in journals.

Summarizing the survey experience, the results provide a comprehensive overview of the participants' experiences and insights regarding the application of DSR in the SE and IS Engineering domain. The responses indicate a broad engagement with DSR methodologies, primarily initiated during PhD studies, driven by both supervisory guidance and the inherent problems the research aimed to solve. Evaluation methods varied, with empirical cycles favouring surveys and focus groups, while engineering cycles leaned towards experiments and interviews. Training and methodological clarity emerged as critical needs, with participants emphasizing the value of early and structured educational support. Despite initial apprehensions, participants found the DSR methodology feasible and ultimately rewarding, with most publishing their findings. Overall, the survey underscores the practical utility and adaptability of DSR, while highlighting areas for enhanced support and training to optimize its application in academic research.

## 7 Conclusion

This work has introduced DSR to identify what aspects must be considered when improving and reinforcing DSR teaching in practical cases. We introduced a set of illustrative examples and discussed them from an educational point of view. Our work also describes the learning objectives of a DSR course, the teaching methodology, thematic units, the teaching plan, before summarising recommendations. Next, we introduced a survey artifact that can be used to gather reflections and experiences of DSR users, especially emerging researchers. The survey incorporates those aspects we consider crucial for a good DSR teaching practice, and it is intended to assist in the ongoing assessment of DSR in SE and IS Engineering to attract expert guidance and interventions required for achieving a sound DSR teaching process. We encourage the community to make use of this survey artifact and contribute, with its adoption, to generate further valuable data for the community to better understand how DSR is used in practice.

Our own survey instance has provided already a useful set of lessons learned that are presented in this chapter. We encourage further adoption and application by the community to gather valuable data that should help fine-tune the DSR teaching process we have proposed. Our own immediate working plan includes conducting a massive data-gathering process intended to collect information on DSR practical use in several universities, oriented towards PhD and MSc students that use Design Science as a research methodology for their research. The survey aims to be naturally improved with the results of such evaluations. Educators may use the survey to assess learners' experience with the DSR application; Moreso, gaps may be identified, which will further inform the DSR learning, design, and materials required for the associated teaching process.

## Online Material

The survey artifact used in this chapter can be found in the online material section

Teaching Design Science as a Method for Effective Research Development

under https://zenodo.org/doi/10.5281/zenodo.11544897.

**Acknowledgments** We would like to acknowledge the participants in the pilot study to assist in evaluating the draft artifact. Their support contributed to the improved artifact which will be used to conduct the main study data from a large sample of participants. This work was developed with the support of the Spanish Ministry of Science and Innovation co-financed by FEDER in the project SREC (PID2021-123824OB-I00).